\documentclass[12pt]{amsart}

\usepackage{amsmath,amssymb}

\makeatletter
\@addtoreset{equation}{section}
\makeatother

\usepackage{enumerate}

\def\ee{\mathrm e}

\def\al{\mathrm {al}}

\def\Sym{\mathrm {Sym}}
\def\wq{ {q}}
\def\SD{\mathrm {SD}}

\def\calE{{\mathcal{E}}}
\def\CC{{\mathbb C}}
\def\RR{{\mathbb R}}
\def\PP{{\mathbb P}}

\newtheorem{definition}{Definition}[section]
\newtheorem{notation}{Notation}[section]
\newtheorem{theorem}{Theorem}[section]
\newtheorem{proposition}{Proposition}[section]
\newtheorem{corollary}{Corollary}[section]
\newtheorem{remark}{Remark}[section]

\def\dfrac#1#2{{\displaystyle\frac{#1}{#2}}}

\def\book#1{\rm{#1}, }
\def\paper#1{\textit{#1}, }
\def\jour#1{\rm{#1}, }
\def\yr#1{({\rm{#1}) }}
\def\vol#1{\textbf{#1}}
\def\pages#1{\rm{#1}}
\def\page#1{\rm{#1}}

\def\publaddr#1{\rm{#1}, }
\def\publ#1{\rm{#1}, }
\def\by#1{{\rm{#1}, }}

\def\eds{\rm{eds.}}

\begin{document}

\title{Relations in a Loop Soliton as a Quantized Elastica}

\author{Shigeki MATSUTANI}

\maketitle

\begin{abstract}
In the previous article (J. Geom. Phys. {\bf 43} (2002) 146),
we show the hyperelliptic solutions of a loop soliton
as a study of a quantized elastica.
This article gives some functional relations
in a loop soliton as a quantized elastica.
\end{abstract}

\bigskip

{\centerline{\textbf{2000 MSC: 37K20, 35Q53, 14H45, 14H70 }}}

\bigskip
\section{ Introduction}

\bigskip

We proposed problems of the quantized elastica in
\cite{Ma7}.
Let a circle $S^1$ immersed in a complex plane $\CC$
 characterized by the affine coordinate
$	Z(s):= X^1(s)+ \sqrt{-1} X^2(s)$ around the origin.
Here $s$ is a parameter of $S^1$
satisfying $ds^2 = (dX^1)^2 + (d X^2)^2$.
For  loops with the Euler-Bernoulli
energy functional,
\begin{gather}
	\calE[Z]= \oint d s \{Z,s\}_{\SD}, \label{1-1}
\end{gather}
where $\{Z,s\}_{\SD}$ is the Schwarz derivative,
\begin{gather}
	\{Z,s\}_{\SD}:=\partial_s\left(\frac{\partial_s^2 Z}
 {\partial_s Z}\right)
       -\frac{1}{2} \left(\frac{\partial_s^2 Z}
             {\partial_s Z}\right)^2
       ,  \label{1-2}
\end{gather}
the quantized elastica problem is to compute
the {\lq\lq}partition function{\rq\rq},
\begin{gather}
	\mathcal Z[\beta]=
\int D Z \exp(-\beta \calE[Z]), \label{1-4}
\end{gather}
in a certain physical sense.
Here $\partial_s := \partial/\partial s$.
Whereas the ordinary (classical) elastica problem is to
 compute its extremal points of the energy functional
 (\ref{1-1}),
in the quantized elastica problem, we should
calculate some contributions from loops with
outside of the extremal points of (\ref{1-1}),
which  contrasts with the
classical elastica problem.

In order to make this physical functional (\ref{1-4})
have mathematical meanings,
we should classify a loop space $\Omega \CC$
of the complex plane with
paying attentions upon the energy functional (\ref{1-1})
and euclidean moves.
In \cite{Ma7}, we studied  the loop space as
the moduli space of the quantized elastica,
\begin{gather*}
 \mathcal M^{\CC}_{\mathrm{elas}}:=
   \{ Z: S^1 \to \CC \ |  \ \oint d Z = 2\pi\}/\sim,
\end{gather*}
where $\sim$ means the euclidean moves.
$\mathcal M^{\CC}_{\mathrm{elas}}$ has a spectrum decomposition,
\begin{gather}
 \mathcal M^{\CC}_{\mathrm{elas}}:= \prod_{E}
  \mathcal M^{\CC}_{\mathrm{elas},E}, \quad
 \mathcal M^{\CC}_{\mathrm{elas}, E}:=
   \{ Z \in\mathcal M^{\CC}_{\mathrm{elas}} \ |\
       \calE[Z] = E \}.
\end{gather}

As the loop soliton, which is defined as follows,
 preserves the local length
and the energy functional  (\ref{1-1}),
\cite{Ma0, Ma7} shows that
 $\mathcal M^{\CC}_{\mathrm{elas}, E}$
consists of the orbits of a group of
its related loop soliton.

\begin{definition}\it
A one parameter family of a loop $\{Z(t)\ | \ t \in \RR\}$
for a real parameter
$t\in \mathbb R$
is called a loop soliton,
if its half curvature
$q:=\dfrac{1}{2\sqrt{-1}} \partial_s \log \partial_s
      Z(t,s)$ obeys the modified Korteweg-de Vries (MKdV)
 equation,
$$
	\partial_t q + 6 q^2 \partial_s q + \partial_s^3 q
	 =0, \label{1-3}
$$
where $\partial_t :=\partial/\partial t$.
\end{definition}

Using the loop soliton, we classified the
loop space $\mathcal M^{\CC}_{\mathrm{elas}}$
in the category of differential
geometry and investigated its topological properties
in \cite{Ma7}.
As a loop soliton is expressed by a hyperelliptic
function of genus $g$ if we include the infinite
genus $g=\infty$, we have an expression
of the partition functions as follows.
\begin{gather}
	\mathcal Z[\beta]=\sum_{g=0}^\infty
		\mathcal Z^{(g)}[\beta], \quad
		\mathcal Z^{(g)}[\beta]:=
\int_{
\mathcal M^{\CC}_{\mathrm{elas},g}}
 D Z \exp(-\beta \calE[Z]), \label{1-4-2}
\end{gather}
where $\mathcal M^{\CC}_{\mathrm{elas},g}$
is a subspace of $\mathcal M^{\CC}_{\mathrm{elas}}$,
whose elements are expressed by hyperelliptic functions of
genus $g$. We give explicit function forms of
loop soliton in terms of Weierstrass hyperelliptic
al functions in \cite{Ma3}.

As J. McKay pointed out, there are apparent
 resemblances between relations in
the replicable functions \cite{FMN, Mc, MS}
and those in the quantized elastica. Thus
we have progressed
 investigations of the resemblances.
In this article, we will study functional
properties of the loop solitons
and a quantized elastica as a sequel of the previous
paper \cite{Ma3}. Our investigations are closely related to
the geometry of MKdV equation studied by Previato in
\cite{P} as mentioned in Remark 3.1 and 3.2.

Section 2 gives minimal preliminaries to express
our results in \S 3 and \S 4.
We start with a hyperelliptic curve given by
(\ref{2-1}) and thus there principally appear no
other parameters  beside $\lambda$'s in (\ref{2-1}).
The quantities defined in Definition
\ref{def-2.3} directly play important roles in our
theory.
After reviewing the previous results \cite{Ma3} in
  Proposition \ref{prop-3.1}, we give
our main theorem in Theorem 3.1.
(\ref{3-3}) is a differential expression as
the generalization to a general genus
of (4.6) in \cite{Ma3}
for the genus one case,
\begin{gather*}
\split
	Z^{(a)}(u) =\lim_{\epsilon\to0} \int^u d u
\frac{1}{\sigma(\epsilon)^2}\exp \Bigr(
-\frac{1}{2}\int^u_\epsilon \int^{u'}_0
\ & \Bigr[ \{Z^{(a)}(u''), u''\}_{\SD} \\
 &-\{Z^{(a)}(u''-\omega_a),u''\}_{\SD}\Bigr] d u'' d u' \Bigr),\\
\endsplit
\end{gather*}
 which  was found
under the stimulus of a formula obtained by J. McKay.
Further we will investigate
the properties of a quantized elastica
based upon the fourier analysis in \S 4
and give our other main
results in Proposition 4.1 and 4.2.

\bigskip

\section{ Preliminary for Hyperelliptic Functions}

\bigskip

\textit{ Hyperelliptic Curve:}
\rm{
This article deals with a hyperelliptic curve $C_g$  of genus $g$
$(g>0)$ given by the affine equation,
\begin{gather} \split
   y^2 &= f(x) \\  &= \lambda_{2g+1} x^{2g+1} +
\lambda_{2g} x^{2g}+\cdots  +\lambda_2 x^2
+\lambda_1 x+\lambda_0  \\
     &=(x-b_1)(x-b_2)\cdots (x-b_{2g+1}),\\
\endsplit  \label{2-1}
\end{gather}
where $\lambda_{2g+1}\equiv1$ and $\lambda_j$'s
and $b_j$'s $(b_i=a_i, b_{g+i}=c_i)$ are  complex numbers.}

\begin{definition} \cite{Ba1, Ba2,  BEL1, W2}

\begin{enumerate}[{(}1{)}]
\item
For a point $(x_i, y_i)\in C_g$,
the unnormalized differentials of the first kind are
defined by,
\begin{gather*}   d u^{(i)}_1 := \frac{ d x_i}{2y_i}, \quad
      d u^{(i)}_2 :=  \frac{x_i d x_i}{2y_i}, \quad \cdots, \quad
     d u^{(i)}_g :=\frac{x_i^{g-1} d x_i}{2 y_i}.
      \label{2-3}
\end{gather*}

\item
The Abel map from $g$-th symmetric product
of the curve $C_g$ to $\CC^g$ is defined by,
\begin{gather*}
u :=(u_1,\cdots,u_g)
:\mathrm{Sym}^g( C_g) \longrightarrow \mathbb C^g,
\end{gather*}
\begin{gather*}
      \left( u_k((x_1,y_1),\cdots,(x_g,y_g)):= \sum_{i=1}^g
       \int_\infty^{(x_i,y_i)} d u^{(i)}_k \right).
      \label{2-6}
\end{gather*}
\end{enumerate}
\end{definition}

\begin{notation}\rm
Let us denote the homology of a hyperelliptic
curve $C_g $ by
$
\mathrm{H}_1(C_g, \mathbb Z)
  =\bigoplus_{j=1}^g\mathbb Z\alpha_{j}
   \oplus\bigoplus_{j=1}^g\mathbb Z\beta_{j}
$.
Here these intersections are given as
$[\alpha_i, \alpha_j]=0$, $[\beta_i, \beta_j]=0$ and
$[\alpha_i, \beta_j]=\delta_{i,j}$.
The complete hyperelliptic integral
of the first kind are defined by,
\begin{gather*}    \pmb{\omega}':=\frac{1}{2}\left[\left(
     \int_{\alpha_{j}}d u^{(a)}_{i}\right)_{ij}\right],
\quad
      \pmb{\omega}'':=\frac{1}{2}\left[\left(
       \int_{\beta_{j}}d u^{(a)}_{i}\right)_{ij}\right],
 \quad
    \pmb{\omega}:=\left[\begin{matrix} \pmb{\omega}' \\ \pmb{\omega}''
     \end{matrix}\right].
  \label{2-9}
\end{gather*}
The Jacobi varieties (Jacobian) $\mathcal J_g$
is defined as a complex torus,
\begin{gather*}
   \mathcal J_g := \mathbb C^g /{\pmb{\Lambda}}_g.
     \label{2-10}
\end{gather*}
Here  $\pmb{\Lambda}_g$   is a real  $2g$-dimensional
lattice generated by the periodic matrix given by $2\pmb{\omega}$.
Further $u$ is the coordinate of $\mathbb C^g$
and of the Jacobian
$\mathcal J_g$.
\end{notation}

\begin{definition}
{\rm{
Using the unnormalized differentials of the second kind,
\begin{gather*}
     d r_{j}^{(i)}=\dfrac{1}{2 y_i}\sum_{k=j}^{2g-j}(k+1-j)
      \lambda_{k+1+j} x_i^k d x_i ,
     \quad (j=1, \cdots, g), \label{2-38}
\end{gather*}
the complete hyperelliptic integral matrices
of the second kind are defined by,
\begin{gather*}    \pmb{\eta}':=\frac{1}{2}\left[\left(
         \int_{\alpha_{j}}d r^{(a)}_{i}\right)_{ij}\right],
\quad
      \pmb{\eta}'':=\frac{1}{2}\left[\left(
        \int_{\beta_{j}}d r^{(a)}_{i}\right)_{ij}\right].
  \label{2-39}
\end{gather*}
The hyperelliptic $\sigma$ function,
which is a holomorphic
function over $u\in \mathbb C^g$, is defined by
[\cite{Ba2}, p.336, p.350], \cite{Kl1, BEL1},
\begin{gather} \sigma(u):=\sigma(u;C_g):
  \equiv\ \gamma\mathrm{exp}(-\dfrac{1}{2}\ ^t\ u
  \pmb{\eta}'{\pmb{\omega}'}^{-1}u)
  \vartheta\negthinspace
  \left[\begin{matrix} \delta'' \\ \delta' \end{matrix}\right]
  (\frac{1}{2}{\pmb{\omega}'}^{-1}u ;\pmb \tau),
     \label{2-40}
\end{gather}
where $\gamma$ is a certain constant factor,
$\vartheta[]$ is the Riemann $\theta$ function,
\begin{gather*}
\vartheta\negthinspace\left[\begin{matrix} a \\ b
 \end{matrix}\right]
     (z; \pmb \tau)
    :=\sum_{n \in \mathbb Z^g} \exp \left[2\pi \sqrt{-1}\left\{
    \dfrac 12 \ ^t\negthinspace (n+a)\pmb \tau(n+a)
    + \ ^t\negthinspace (n+a)(z+b)\right\}\right],
     \label{2-41}
\end{gather*}
with $\pmb \tau:={\pmb{\omega}'}^{-1}\pmb{\omega}''$
for $g$-dimensional vectors $a$ and $b$,
and
\begin{gather*}
 \delta' :=\ ^t\left[\begin{matrix} \dfrac {g}{2} & \dfrac{g-1}{2}
       & \cdots
      & \dfrac {1}{2}\end{matrix}\right],
   \quad \delta'':=\ ^t\left[\begin{matrix} \dfrac{1}{2} & \cdots
& \dfrac{1}{2}
   \end{matrix}\right].
     \label{2-42}
\end{gather*}
}}
\end{definition}

\bigskip

\bigskip
\begin{definition}\label{def-2.3}
\rm{
\begin{enumerate}[{(}1{)}]
\item Hyperelliptic $al$ function is defined by
[\cite{Ba2} p.340, \cite{W2}],
\begin{gather}
\mathrm{al}_r(u) = \gamma_r\sqrt{F(b_r)} , \label{2-20}
\end{gather}
where $\gamma_r:=\sqrt{-1/P'(b_r)}$  and
\begin{gather}
	F(x):= (x-x_1) \cdots (x-x_g).
          \label{2-21}
\end{gather}

\item
Hyperelliptic $\zeta_{\nu}$ function is defined by,
\begin{gather}
	\zeta_{\mu}=\frac{\partial}{\partial u_\mu}
           \log \sigma(u).
  \label{2-44-1}
\end{gather}

\item
Hyperelliptic $\wp_{\mu\nu}$ function is defined by,
\begin{gather*}
	\wp_{\mu\nu}=-\frac{\partial^2}
          {\partial u_\mu \partial u_\nu}
           \log \sigma(u).
  \label{2-44-2}
\end{gather*}

\item
The power symmetric function $\wq$ is defined by
\begin{gather}
	{\wq}_{n}:=\sum_{i=1}^g x_i^n(u), \quad
	{\wq}_{n, \mu } :=
\frac{\partial}{\partial u_\mu} \wq_n.
  \label{2-45}
\end{gather}

\end{enumerate}
}
\end{definition}

On the choice of $\gamma_r$,
we will employ the convention of Baker \cite{Ba2}
instead of original
one \cite{ W2}.

\bigskip

\begin{proposition}
\begin{enumerate}
\item
Introducing the half-period $\omega_r :=
\int^{b_r}_\infty du^{(a)}$,
we have the relation {\rm[\cite{Ba2}, p.340]},
\begin{gather}
	\al_r(u) =\gamma_r'' \frac{
\exp(-{}^t u  \pmb{\eta}'{\pmb{\omega}'}^{-1}\omega_r)
\sigma( u + \omega_r)}{\sigma(u)},
           \label{2-43}
\end{gather}
where $\gamma_r''$ is a certain constant.

\item
The hyperelliptic $\wp_{g i}$ function is given as
an elementary symmetric function,
\begin{gather*}
F(x)=x^g-\sum_{i=1}^g \wp_{g,i} x^{g-i}.\label{2-22}
\end{gather*}
\textit{i.e.,}
\begin{gather}
	\wp_{g\nu}=(-1)^i e_{\mu-1}(u),
  \label{2-44-3}
\end{gather}
where $ e_{\mu}(u)$ is the $\mu$-th elementary
symmetric function
of $x_i$'s.

\end{enumerate}
\end{proposition}

\bigskip
\section{Relations in a loop soliton}

As mentioned in Introduction,
this section gives  relations in a quantized elastica
following the previous results.
Before we will show our new results,
we review the previous results
 in \cite{Ma3} as follows.

\begin{proposition}\label{prop-3.1}
Let the configuration of the $x$-components
$(x_1, \cdots, x_g)$ of the affine coordinates
of the hyperelliptic curves $\Sym^g(C_g)$ and
the coefficients $\lambda$'s of each $C_g$
satisfy,
\begin{gather}
	| F(b_r) | =1, \quad \mbox{and},\quad
      u_g \in \RR. \label{3-7}
\end{gather}
For such $(x_1,y_1),\cdots,(x_g,y_g)$, we have
$\mathbf u:=\mathbf u( (x_1,y_1),$ $\cdots,(x_g,y_g) )$
and following results.
\begin{enumerate}

\item By setting $s\equiv  u_g$ and $t\equiv u_{g-1} +
( \lambda_{2g-1} +b_r) u_g$,
\begin{gather*}
   \partial_{u_g}Z^{(r)} :=  F(b_r), \quad
\text{or} \quad |\partial_{u_g} Z^{(r)} | = 1,  \label{3-8}
\end{gather*}
completely characterizes the loop soliton.

\item The shape of loop soliton is given by,
\begin{gather*}
      Z^{(r)}= \frac{1}{r_0}\left(b_r^g u_g + \sum_{i=1}^g
                b_r^{i-1} \zeta_{i}\right). \label{3-9}
\end{gather*}

\item
The Schwarz derivative of $Z$ with respect to $u_g$
\begin{gather}
 \{Z^{(r)},u_g\}_{\SD}=4\wp_{g g}+2\lambda_{2g}+2b_r.
       \label{3-18}
\end{gather}

\end{enumerate}

\end{proposition}

\bigskip
Here we should give remarks on the previous results.

\begin{remark}{\rm{
\begin{enumerate}
\item
Though we did not notice neither
mention in \cite{Ma3, Ma7}, we must say that
parts of the results in \cite{Ma3, Ma7} had
been already obtained in \cite{P} using
Riemann $\theta$ functions.

\item
In order to satisfy (\ref{3-7}) for real parts
in the Jacobian,
we must constraint the coefficients in (\ref{2-1}) of
the curve. However we could not find such conditions in
this stage like genus one case in \cite{Mu}.
Thus we plan to  consider these conditions.

\end{enumerate}
}}

\end{remark}

\bigskip

From the previous results, we
automatically have following corollary.
\begin{corollary}
A loop soliton $Z^{(r)}(u)$ satisfies
following relations,
\begin{enumerate}

\item
\begin{gather}
 \partial_{u_g} Z^{(r)}(u+2\omega_i')
       \equiv \partial_{u_g}Z^{(r)}(u),
\quad
\partial_{u_g}Z^{(r)}(u+2\omega_i'')
        \equiv \partial_{u_g}Z^{(r)}(u).
      \label{3-1-1}
\end{gather}

\item When we regard $Z^{(r)}$ as a function of
$(x_i-b_r)_{i=1,\cdots,g}$,
\begin{gather}
\overline{\partial_{u_g}
Z^{(r)}(x_1 - b_r,  \cdots, x_g-b_r)}
\equiv \partial_{u_g} Z^{(r)}
\left(\frac{1}{x_1 - b_r}, 
       \cdots, \frac{1}{x_g - b_r}\right).
      \label{3-1-2}
\end{gather}

\end{enumerate}
\end{corollary}

Followings are our main results in this article.
\begin{theorem}
A loop soliton $Z^{(r)}(u)$ satisfies
following relations,
\begin{enumerate}

\item
\begin{gather*}
 \partial_{u_g} Z^{(r)}(u)=
         b_r^g \exp
\left( - \sum_{n=1}^\infty \frac{q_n}{n}b_r^{-n} \right)
      \label{3-1-3}
\end{gather*}

\item
\begin{gather}
 \{Z^{(r)}(u+\omega_r), u_g\}_{\SD}+\{Z^{(r)}(u), u_g\}_{\SD}
= -\sum_{n, m = 1}^\infty \frac{\wq_{n,g}
      \wq_{m,g}}{n m } b_r^{-n-m}.
      \label{3-2}
\end{gather}

\item
\begin{gather}
\frac{1}{2}\left[
 \{Z^{(r)}(u+\omega_r), u_g\}_{\SD}-\{Z^{(r)}(u), u_g\}_{\SD}
\right]
= -\partial_{u_g}^2
\log\left(\partial_{u_g} Z^{(r)}(u) \right).
      \label{3-3}
\end{gather}

\end{enumerate}
\end{theorem}

\bigskip

\begin{proof}
The first formula is obvious from the relation between $F(b_r)$ and
the power symmetric functions $q_n$.
Due to (\ref{2-43}), we have
\begin{gather*}
-\partial_{u_g}^2
\log(F(b_r)) = - 2 \wp_{g g}(u+w_r) + 2 \wp_{g g}(u).
\end{gather*}
(\ref{3-18}) leads us the third formula (\ref{3-3}).

From (3.27) in \cite{Ma4}, which is essentially
the Miura transformation, we have
\begin{gather*}
-\partial_{u_g}^2
\log(F(b_r)) =  4 \wp_{g g}(u) + 2 \lambda_{2g}
        + 2 b_r + \frac{1}{2}\left(
\partial_{u_g}
\log(F(b_r))\right)^2.
\end{gather*}
As mentioned in (3.8) of \cite{Ma4},
\begin{gather*}
	\frac{\partial}{\partial u_g }
         =\sum_{i=1}^g\frac{2y_i}{F'(x_i)}
             \frac{\partial}{\partial x_i} ,
\end{gather*}
we have
\begin{gather*}
\partial_{u_g}
\log(F(b_r))=
\sum_{n = 1}^\infty \frac{q_{n,g}}{n } b_r^{-n}.
\end{gather*}
These constitute the second formula (\ref{3-2}).
\end{proof}

\bigskip

\begin{remark}{\rm{
\begin{enumerate}
\item
(\ref{3-3}) is the generalization of (4.6)
in \cite{Ma3}, which is
the same formula of genus one, to
s general genus $g$.

\item
$F(b_r)$ can be regarded as
a generation function of the elementary
symmetric functions
and thus behind our theorem, the Newton formula plays
important roles.

\item
It is noted that
$F(x)$ and $\partial_{u_g}F(x)$ appeared in
the book of Mumford \cite{Mu0} as $U(x)$ and $V(x)$
in his triplet representation $(U, V, W)$
of functions of hyperelliptic curves.

\item
(\ref{3-1-1}) can be a stronger relation for a
closed loop soliton,
{\it{ i.e.}},
\begin{gather}
 Z^{(r)}(u_1, \cdots, u_{g-1}, u_g + 1)
=Z^{(r)}(u_1, \cdots, u_{g-1}, u_g).
      \label{3-6-1}
\end{gather}
In this case, we have fourier expansion as shown in
next section.

\item On a loop soliton and geometry of MKdV equations,
readers should consult the reference \cite{P},
which gives several mathematical open problems and
results in issues related to our quantized elastica
problem.

\item As in remarked in (\cite{P}, 4.3),
our system is closely related to the
 formula 27 in p.19 of \cite{F},
which is of the Schwarz derivative
and prime form.
The $\al$-functions are solutions of
the Dirac equation in \cite{Ma3},
which is the spinor representation
of the Frenet-Serret equation.
We should connect them in future.

\end{enumerate}

}}
\end{remark}

\bigskip
\section{Winding loops}

Due to the (\ref{3-6-1}), we have fourier expansions
of $Z^{(r)}$ of a closed loop soliton or a quantized elastica,
{\it i.e.},
using functions $a_n$ of $(u_1,u_1, \cdots, u_{g-1})$
and real parameter $s$,
\begin{gather*}
 Z^{(r)}(u) = \sum_{n=-\infty}^\infty \frac{1}{\sqrt{2\pi}}
a_n
\ee^{2\pi\sqrt{-1}n s },
\quad
\frac{1}{\sqrt{-1}}\partial_{s} Z^{(r)}(u) =
 \sum_{n=-\infty}^\infty
\sqrt{2\pi} n a_n
\ee^{2\pi\sqrt{-1}n s}.
      \label{3-6-2}
\end{gather*}
In this sense, we will regard $Z^{(r)}(u)$ as
a function of $s$ with parameters  $u^\#:=(u_1,u_1, \cdots, u_{g-1})$
and refer it by $Z^{(r)}(s):=Z^{(r)}(u^\#;s)$.
Then we have the following proposition.

\begin{proposition}
\begin{enumerate}
\item The euclidean move is represented by
a choice of $a_0$ and global constant factor $c$ of
$Z^{(r)}$.

\item In terms of the real parameter $s$, there exists
a complex number $c$,
\begin{gather*}
       \overline{Z^{(r)}(u^\#;s)-a_0} = c (Z^{(r)}(u^\#;-s) -a_0), \qquad
       \overline{a_n} =c a_{n}, \quad\mathrm{ for }\ n\neq0.
\end{gather*}

\item By choosing $a_0=0$ and $c=1$,
the reality condition
$|\partial_s Z(u^\#;s)|=1$ is expressed by
\begin{gather*}
 2\pi\sum_{m=-\infty}^\infty n (n+m) a_{m} a_{n + m} = \delta_{n,0}.
   \label{3-6-6}
\end{gather*}

\item For $a_0=0$ and $c=1$,
the fourier coefficients of the curvature of $Z^{(r)}(-s)$
can be expressed by the bilinear form of $a_n$'s,
\begin{gather*}
\frac{1}{\sqrt{-1}}\partial_{s} \log\partial_{s}
 Z^{(r)}(u^\#;s) =
- \sum_{n=-\infty}^\infty
\left( 4\pi^2\sum_{m=-\infty}^\infty
(n+m)^2(m) a_m a_{n+m}\right)
\ee^{2\pi\sqrt{-1}n s}.
      \label{3-6-7}
\end{gather*}

\end{enumerate}
\end{proposition}

\begin{proof}
The first relation is obvious. The complex conjugate determines the
orientation of the complex plane while
that of a loop  is by the orientation of the arclength
parameter $s$. Thus the second relation is justified.
Noting
$\overline{\partial_{s}Z^{(r)}}=1/\partial_{s}Z^{(r)}$,
direct computations give the third and fourth relations.
\end{proof}

As a loop soliton and an element in
$\mathcal M^{\CC}_{\mathrm{elas}}$
 defined as a loop
 modulo euclidean moves,
we should regard $Z^{(r)}$ as a vector in the
complex plane. It implies that an addition of
$Z^{(r)}$'s with complex coefficients
has mathematical meanings.

Further as mentioned in \cite{Ma1},
there are winding solutions in our moduli space
$\mathcal M^{\CC}_{\mathrm{elas}}$.
Hence we will define a winding loop soliton
for a loop soliton $Z^{(r)}(u^\#;s)$,
\begin{gather}
Z^{(r,n)}(u^\#;s):=\frac{1}{n}Z^{(r)}(u^\#;n s).
\end{gather}

The winding loop solitons has  following
properties, which are not difficult to be proved.

\begin{proposition}

For a natural number $n$ and
a prime number $p$, we have
following relations,
\begin{enumerate}

\item
\begin{gather}
\calE[Z^{(r,n)}]=n^2\calE[Z^{(r)}].
\end{gather}

\item
\begin{gather*}
 p Z^{(r, p n)}(u^\#;s) = Z^{(r, n)}(u^\#; p s).
\end{gather*}
\item
\begin{gather}
  \left(Z^{(r,p n)}\left(\frac{s}{p}\right)
             +Z^{(r,p n)}\left(\frac{s+1}{p}\right)+\cdots
             +Z^{(r,p n)}\left(\frac{s+p-1}{p}\right)
              \right) = Z^{(r, n)}(s).
         \label{eq:4-10}
\end{gather}

\end{enumerate}
\end{proposition}

\bigskip

\begin{remark}{\rm{
\begin{enumerate}
\item
The relation (\ref{eq:4-10}) remind of the action of Hecke for
modular function of vanishing
weight and for a prime number $p$ \cite{S},
\begin{gather}
p T_p (f(z)) = f(pz) +\left(f\left(\frac{z}{p}\right)
             +f\left(\frac{z+1}{p}\right)+\cdots
             +f\left(\frac{z+p-1}{p}\right)
              \right) .
\end{gather}

\item
We should note the partition function (\ref{1-4}).
Even though (\ref{1-4})
could not computed in this stage, we can
compute its part, $\mathcal Z^{(a)}[\beta]$
($a=1,2$).
We know the closed loop soliton
solutions of genera zero and one explicitly,
which is given by disjoint types, {\it i.e.},
a circle and an eight-figure shape \cite{Ma0}.
Considering contributions of winding loop soliton,
 for $a=0,1$, we obtain,
\begin{gather*}
\mathcal Z^{(a)}[\beta] = \sum_{n=1}^\infty \ee^{-\beta n^2 E_a}
        = \frac{1}{2}
    \left(\theta( \sqrt{-1} \beta E_a/\pi) - 1\right),
\end{gather*}
where $E_0$ and $E_1$ are the energies of genera
zero and one and  $\theta(z) $ is the elliptic theta function,
$\theta(z) := \sum_{n=-\infty}^\infty \ee^{\sqrt{-1}\pi z n^2 }$.
Due to properties of the elliptic theta function
and Poisson sum formula,
\begin{gather*}
\mathcal Z^{(a)}\left[\beta\right]
 = \sqrt{\frac{1}{E_a\beta} }\sum_{n=1}^\infty \ee^{- n^2 /E_a\beta} +
\frac{1}{2}\left(\frac{1}{\sqrt{E_a\beta}}-1\right).
\end{gather*}
As $\mathcal Z^{(a)}\left[\beta+2\pi\sqrt{-1}/E_a\right]$
$=\mathcal Z^{(a)}\left[\beta\right]$, we regard that
$\mathcal Z^{(a)}\left[\beta\right]$ has modular properties.

When we approximate $\mathcal Z[\beta]$ by
$\mathcal Z^{(a)}\left[\beta\right]$ or
$\mathcal Z^{(0,1)}[\beta]:=\sum_{a=0}^1 \mathcal Z^{(a)}[\beta]$,
we might encounter a critical phenomena
from the viewpoint of statistical physics due to the
modular properties.
\end{enumerate}

}}
\end{remark}

\section{Acknowledgment}

We thank Prof. E.~Previato,  Prof. J. McKay and Prof. Y. \^Onishi
 for helpful suggestions  and
encouragements.


\bigskip

\bigskip

{Shigeki Matsutani}

{e-mail:RXB01142\@nifty.com}

{8-21-1 Higashi-Linkan}

{Sagamihara 228-0811 Japan}

\end{document}